\documentclass[review,number,sort&compress]{elsarticle}

\usepackage{amsmath}
\usepackage{subcaption}
\usepackage{longtable}
\usepackage{pdflscape} 

\journal{Nucl. Instr. Meth. Phys. Res. - A}

\begin{document}

\begin{frontmatter}

\title{Testing FLUKA on neutron activation of Si and Ge at nuclear research reactor using gamma spectroscopy
}


\author[pucp]{J. ~Bazo}
\ead{jbazo@pucp.edu.pe}
\author[pucp]{J.M. ~Rojas}
\author[pucp]{S. ~Best}
\author[ipen1]{R. ~Bruna}
\author[pucp]{E. ~Endress}
\author[ipen2]{P. ~Mendoza}
\author[ipen2]{V. ~Poma}
\author[pucp]{A.M. ~Gago}

\cortext[cor1]{Corresponding author}
\address[pucp]{Secci\'on F\'isica, Departamento de Ciencias, Pontificia Universidad Cat\'olica del Per\'u, Av. Universitaria 1801, Lima 32, Per\'u}
\address[ipen1]{C\'alculo An\'alisis y Seguridad (CASE), Instituto Peruano de Energ\'ia Nuclear (IPEN), Av. Canad\'a 1470, Lima 41, Per\'u}
\address[ipen2]{Divisi\'on de T\'ecnicas Anal\'iticas Nucleares, Instituto Peruano de Energ\'ia Nuclear (IPEN), Av. Canad\'a 1470, Lima 41, Per\'u}

\begin{abstract}
Samples of two characteristic semiconductor sensor materials, silicon and germanium, have been irradiated with neutrons produced at the RP-10 Nuclear Research Reactor at 4.5 MW. Their radionuclides photon spectra have been measured with high resolution gamma spectroscopy, quantifying four radioisotopes  ($^{28}$Al, $^{29}$Al for Si and $^{75}$Ge and $^{77}$Ge for Ge). We have compared the radionuclides production and their emission spectrum data with Monte Carlo simulation results from FLUKA. Thus we have tested FLUKA's low energy neutron library (ENDF/B-VIIR) and decay photon scoring with respect to the activation of these semiconductors. We conclude that FLUKA is capable of predicting relative photon peak amplitudes, with gamma intensities greater than 1\%, of produced radionuclides with an average uncertainty of 13\%. This work allows us to estimate the corresponding systematic error on neutron activation simulation studies of these sensor materials. 

\end{abstract}


\end{frontmatter}


\section{Introduction}

At nuclear research facilities semiconductor materials are irradiated with neutrons in order to obtain transmutation doping with high homogeneity. This irradiation method is also used to reproduce radiation damage of active sensors in the context of LHC experiments, among other aims. For instance, there are several studies on neutron irradiation of silicon \cite{rad_meas_Si_n, nIrrad_Si_detectors, irrad_Si_detectors, ATLAS_HL-LHC} and germanium \cite{damage_Ge2,damage_Ge,damage_Ge3} detectors. 

Neutron irradiation damage is the consequence of atomic displacements and non-elastic nuclear reactions such as He/H gas production and also solid nuclear transmutation \cite{damage_type}. For example, the atomic displacement alters the microstructure (generating net defects), while the nuclear transmutation changes the chemical composition of materials, modifying in general their properties, such as thermal conductivity and electrical resistivity. In this work, we will focus on the irradiation effects related to nuclear transmutation comparing its measurement with Monte Carlo simulation.  

We irradiate different samples of two characteristic semiconductor sensor materials, silicon and germanium, with the wide spectrum neutron flux produced at the core of the RP-10 \cite{RP10_IPEN} Nuclear Research Reactor at IPEN (Peruvian Institute of Nuclear Energy). Then, we measure their associated radionuclides photon spectra with high resolution gamma spectroscopy.

The Monte Carlo simulation has been performed using FLUKA (FLUktuierende KAskade) \cite{ FLUKA1, FLUKA2}. This simulation package describes the radiation interaction and transport in detector materials. In the context of this work, we use FLUKA to estimate the radionuclides production and their emitted gamma spectrum for the neutron irradiation tests already mentioned.

We have carried out a comparison of the experimental data against the FLUKA simulation. A similar FLUKA validation study, \cite{validation_FLUKA}, irradiated construction materials of high-energy accelerators using the radiation field of a 120 GeV positive hadron beam stopped in a copper target. Thus they explored a higher energy regime and different materials. Our results are in the energy range below 10 MeV and are useful for understanding the accuracy of FLUKA in simulating nuclear solid transmutation as a consequence of neutron irradiation. This serves as benchmark of the low energy neutron library (ENDF/B-VIIR) and decay photon scoring. 

This paper is divided as follows: we first review the theoretical aspects of the neutron activation analysis. Then, we describe in detail the measurements performed at the nuclear research reactor. Next, we outline the simulation implemented in order to compare it, in the results section, to the experimental data. We conclude estimating the accuracy of FLUKA in predicting the activation products gamma spectrum. 

\section{Neutron Activation Analysis}
\label{Sec:neu_act}

When a sample is exposed to a high neutron flux in a nuclear reactor core, artificial radionuclides are produced. For instance, in a process referred to as neutron activation, a stable nucleus can absorb (i.e. capture) a free neutron, resulting in an unstable isotope state. The unstable isotope can undergo some de-excitation process reaching a stable nuclide, leading in most cases to observable gamma peaks \cite{ENSDF}.

A nuclear thermal reactor, as the one we have used, is characterized by its neutron energy spectrum which has three well defined regions: thermal, epithermal and fast neutrons. The neutron activation cross section is most relevant in the thermal regime and also for fast neutrons. The corresponding most likely reactions are the capture of thermal neutrons (n,$\gamma$) and the capture of fast neutrons (n, p) or (n,$\alpha$).

Other reactions such as (n,2n), (n,np) and (n,d) are not energetically possible below $\approx$10 MeV, i.e. the maximum energy of a thermal nuclear research reactor. The neutron activation cross section \cite{Sigma} for each process depends on the target isotope and neutron energy. 

The process to follow in order to measure the  gamma emission from radionuclides starts by irradiating the sample during a time period, {\it irradiation time}. Then the gamma spectroscopy measurements begin. During the elapsed time between the end of irradiation and the initial measurement, {\it decay time}, the radionuclides can decay. Next, the sample is measured in the detector during an interval, {\it measure time} ($t_m$), when the decaying isotopes are counted. Both irradiation and decay times are input in the FLUKA simulation code.

On the other side, to estimate the corrected total number of emitted photons, from the detected ones, the following factors have to be considered: the detector efficiency $\epsilon$, that depends on energy and distance of the source to the detector and the detector's deadtime, $t_{dead}$. Then the corrected counting rate $N$, given the measured counting rate $M$, is:
\begin{equation}
N=\frac{M}{\epsilon}\left(\frac{t_{m}}{t_{m}-t_{dead}}\right)
\end{equation}

\section{Neutron irradiation measurements}
\label{Sec:neu_irr_measur}

We have irradiated one silicon and two germanium samples followed by measurements with a gamma spectrometer.

\subsection{Nuclear Reactor description}
\label{subsec:nuc_reactor}
The neutron source is the RP-10 \cite{RP10_IPEN} Nuclear Research Reactor at IPEN (Peruvian Institute of Nuclear Energy) in Huarangal, near Lima, Peru. The RP-10 is a MTR (Material Testing Reactor) pool-type reactor with U$_3$O$_8$ nuclear fuel containing 19.75\% enrichment in $^{235}$U. The core is located at the bottom of a cylindrical tank (4m diameter and 11m deep) and is surrounded by graphite  and beryllium neutron reflectors. The tests were carried out using the same power at 4.5 MW and exposure times of 20 s and 60 s. 

\begin{figure}[ht]
\centering
  \includegraphics[width=0.5\textwidth]{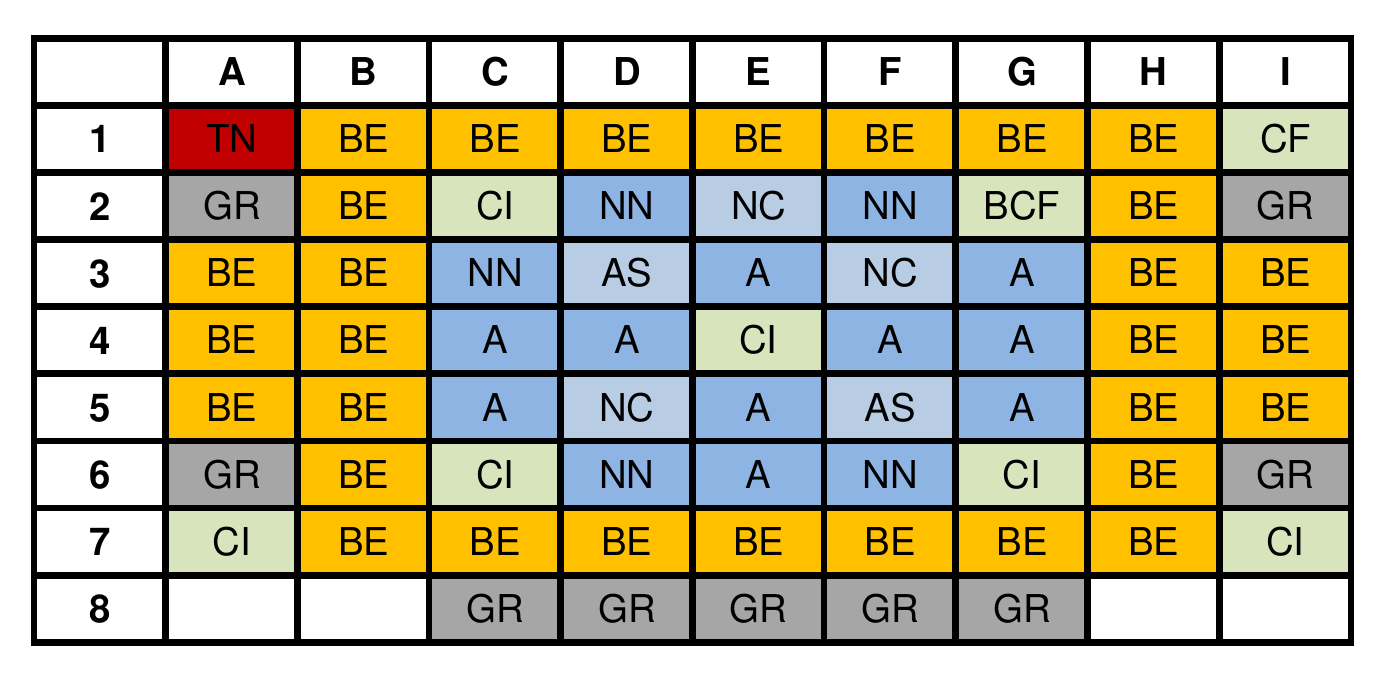}
  \caption{Configuration scheme of RP-10's 44 core. The codes are: TN (used irradiation position), A and NN (nuclear fuel element), AS and NC (control fuel element), BE(beryllium reflector), GR (graphite), CI (irradiation box), BCF (fine control bar) and CF (fission counter).}
  \label{fig:RP10_core}
\end{figure}	

\begin{figure}[ht]
\centering
  \includegraphics[width=0.5\textwidth]{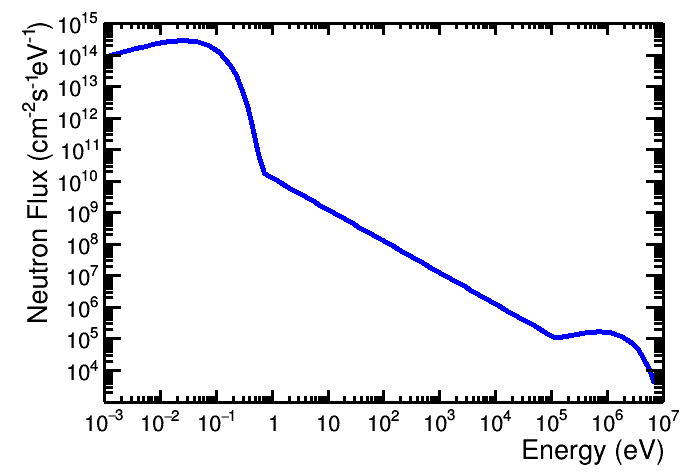}
  \caption{Theoretical neutron spectrum of RP-10 nuclear reactor at position A1 normalized with experimental integrated fluxes obtained with the activated foil dosimetry $k_{0}$ parametric method \cite{Pablo}. This spectrum is used in the FLUKA simulation.}
  \label{fig:RP10_neutron_spectrum}
\end{figure}

The nuclear core configuration number 44 (see Fig. \ref{fig:RP10_core}) was used. The nominal neutron fluxes measured at 4.5 MW in the irradiation position A1 (corner of the core grid) have been estimated using the activated foil dosimetry $k_{0}$ parametric method described in \cite{Pablo}. These experimentally estimated neutron fluxes are given in Table \ref{tab:Norm_nfluxes} for different energy ranges with their respective errors. The overall error, taking into account the uncertainty in the spectrum shape, is $\approx$10\%. The theoretical neutron spectrum of a thermal reactor, shown in Fig. \ref{fig:RP10_neutron_spectrum}, was used in the FLUKA simulation. It has been normalized in each energy range with the energy-integrated measured flux given in Table \ref{tab:Norm_nfluxes}. The explicit flux (cm$^{-2}$s$^{-1}$eV$^{-1}$) formulas, which are continuous in their limiting energies, are given below: 

\begin{equation}
\begin{split}
\mbox{Thermal}~[0~\mbox{eV}, 0.625~\mbox{eV}] &=2.8554\times 10^{15} \sqrt{E}  e^{-18.22 E}   \\
\mbox{Epithermal}~[0.625~\mbox{eV}, 122~\mbox{keV}]&=1.2642\times 10^{10} \left( \frac{1}{E} \right)   \\
\mbox{Fast}~[122~\mbox{keV}, 10~\mbox{MeV}]&=211833 \sinh{\sqrt{2.29\times 10^{-6} E}} e^{-1.036\times 10^{-6} E} 
\end{split}
\end{equation}

\begin{table}[ht]
\centering
\begin{tabular}{|c|c|c|c|}
\hline 
Measured flux&Thermal & Epithermal & Fast \\ 
& [0 eV,0.625 eV] & [0.625 eV,122 keV] & [122 keV,10 MeV] \\ 
\hline 
 $\times10^{11}$(n/cm$^2$s)&$86\pm1.72$ & $1.54\pm0.15  $ & $4.59 \pm0.05$ \\ 
\hline 
\end{tabular} 
\caption{Neutron fluxes in different energy ranges measured at position A1 with 4.5 MW using activated foil dosimetry\cite{Pablo}.}
 \label{tab:Norm_nfluxes}
\end{table}

Each sample was stacked into a polyethylene irradiation container. The containers were sent for irradiation using the pneumatic transfer system to the A1 position  of the RP-10. Since the irradiation position is at the corner of the core, were the reflectors are located, the flux is no longer isotropic as in the center. A possible error due to this positioning is already taken into account since the neutron flux has been measured at this precise location.

\subsection{Samples description}
The silicon and germanium samples are pieces of single side polished (SSP) wafers, 0.5 mm thick, undoped, with $<100>$ orientation. Their properties (element, mass), irradiation time and detection specifications (distance to detector, decay time, measurement time and dead time) are summarized in Table~\ref{tab:samples_description} showing, for each sample, four measurements at different times. The whole set of measurements is used when finding the global normalization described in Sec. \ref{Sec:results}. 

The chemical composition of the samples was studied previously using SEM-EDX (Scanning Electron Microscope / Energy-Dispersive X-ray) spectroscopy (models FEI Quanta
650 and EDAX TEAM$^{\mathrm{TM}}$ EDS) with a detection limit of 0.01\%. Both silicon and germanium samples were found to have a high purity $100^{+0.0}_{-0.77}$\%. In the case of silicon, primary line K and secondary line L were identified and for germanium primary lines K$\alpha$ and K$\beta$, as well as secondary line L were singled out. 

\begin{table}[ht]
\centering
\begin{tabular}{|c|cccccc|}
\hline 
Element & Mass & Irradiation & Distance & Decay & Measure & Dead \\ 
 &  (mg) & time (s)& (mm)& time (s) & time (s) & time (\%)  \\ 
\hline 
Si & 41.38 & 60 &119 & 202 & 168.33 &7.9\\ 
  & & & & & 316.27& 5.9\\ 
  & & & & & 664.6 & 3.6\\ 
  & & & & & 1210.9 & 2.2\\ 

\hline 
Ge & 9.48 & 20 &239 & 235 & 205.53& 2.7 \\ 
  & & & & & 351.1 & 2.2\\ 
  & & & & & 558.8 & 1.9\\ 
  & & & & & 991.7 & 1.7\\ 

\hline 
Ge & 11.87 & 20 &239 & 237 & 125.61 & 3 \\ 
  & & & & & 269.28 & 2.3\\ 
  & & & & & 608 & 1.9\\ 
  & & & & & 991.7 & 1.7\\ 

\hline 
\end{tabular} 
\caption{Samples irradiation and detection parameters.  Measurements at different times for each sample and their corresponding dead times are given.}
\label{tab:samples_description}
\end{table}

\subsection{Gamma spectrometer description}

We carried out the gamma spectrum measurements with a High Purity (HP) Ge semiconductor detector: ORTEC model GEM70P4 (70\% relative efficiency and 1.9 keV resolution (FWHM) for the 1332.5 keV peak of $^{60}$Co and 1 keV resolution at 122 keV). For each sample several gamma spectroscopy measurements were performed at different times. Measurements were done using a 3D-printer sample holder with custom made geometry such as to minimize the gamma attenuation. 

\begin{figure}[ht]
\centering
  \includegraphics[width=0.5\textwidth]{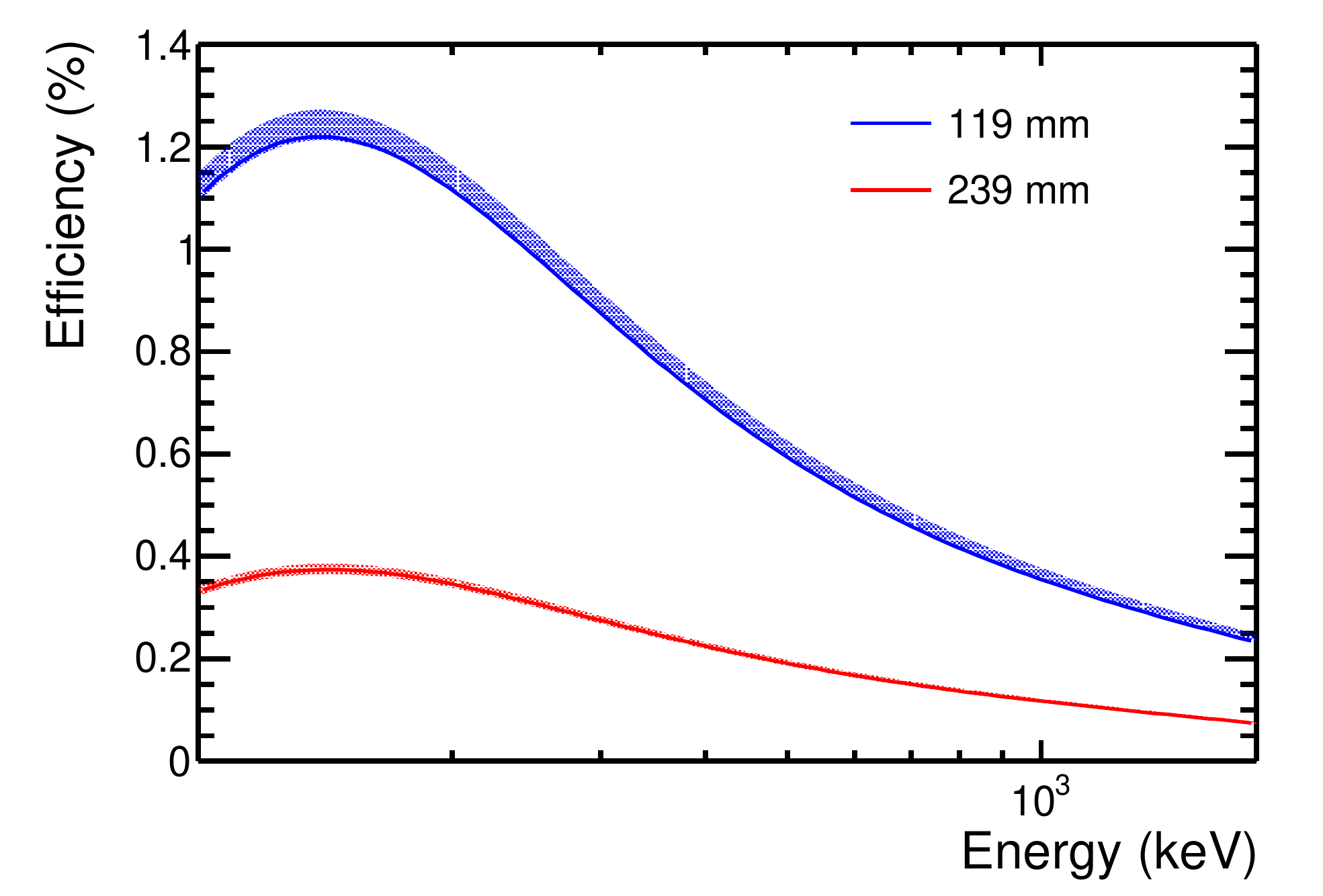}
  \caption{Efficiency of ORTEC model GEM70P4 spectrometer used at two different detection distances. The shadowed bands represent the estimated error.}
  \label{fig:Spect_eff}
\end{figure}

The efficiency of the spectrometer, including error bands, as a function of the energy at the two measurement distances are shown in Fig.\ref{fig:Spect_eff}. The function is a 4-degree polynomial fit to the efficiency data. The efficiency calibration has been performed from 100 keV to 1800 keV, thus energies outside this range have a much higher uncertainty. This detection efficiency correction is applied afterwards in order to account for the gamma spectroscopic measurement specific characteristics, not described in the simulation.

Under the conditions described above, the peaks and radionuclides, product of the nuclear reactions and decays, were determined. We have worked with a deadtime lower than 8\% and 3\%, for silicon and germanium respectively, in order to maintain the pulse shaping formation. However, as a side effect, we get lower statistics for some gamma peaks that are later removed, as mentioned in Sec. \ref{Sec:results}.
 
\section{FLUKA simulation implementation}

We have performed FLUKA simulations for a simplified description of the experiment. The samples are exposed to a direct and collimated neutron beam that follows the spectrum of the RP-10 reactor. Then the photon spectrum emitted by the radioactive nuclides is recorded.

A user-defined source routine was implemented for the neutron energy spectrum of the RP-10 from tabulated data of the flux described in Sec. \ref{subsec:nuc_reactor}. In order to represent the irradiation, a 1cm-radius circular neutron beam is used pointing towards the center of the sample, hitting it perpendicularly to its longitudinal section. The sample geometry in simulation is taken as a right circular cylinder (RCC) of 1 cm radius and 50$\times10^{-4}$ cm height for simplicity, since the actual shape of the real samples were not entirely uniform. 

The simulation defaults have been set to {\it{precision}} activating the corresponding flag. In addition, the simulation of radioactive decays (RADDECAY) has been activated. The irradiation profile (IRPROFI) for radioactive decays uses the irradiation intervals given in Table \ref{tab:samples_description} with an equivalent beam intensity of $\approx$2.9$\times10^{13}$ neutrons/s (total energy integrated flux in the beam area). A related point to consider is that FLUKA cannot fully simulate the transmutation of the whole material, since it keeps constant the target material composition throughout the entire simulation run. However, given the size of the sample and the irradiation time and flux, it is not an important issue. 

In order to improve statistics, each radioactive nucleus is {\it{decayed}} in 10 replicas. We used the activation study mode where the time evolution is calculated analytically and all daughter nuclei and associated radiation are considered at fixed experimental decay times (DCYTIMES) given in Table \ref{tab:samples_description}. At these times after irradiation, decay photons from radioactive products are scored (DCYSCORE) over a spherical region surrounding the sample. The measurement duration is introduced analytically after simulation using Eq.\ref{Eq_timeevol}.

In addition, residual nuclei, produced in the inelastic interactions inside the nuclear reactor, have been scored (RESNUCLEI) in the target (sample) region in order to compare them with the expected radionuclides to be observed. 

Independent simulations were performed with 10$^{10}$ and 10$^9$ primary particles for silicon and germanium, respectively.

\section{Validation results}
\label{Sec:results}
The gamma spectrometer measurements of the samples were performed in the energy range from 39 keV to 3223 keV. The nuclide and peak analysis report is given in Table \ref{tab:validation_results_Si_Ge}. 

In general, silicon has three natural occurring isotopes: $^{28}$Si, $^{29}$Si and $^{30}$Si, with average relative isotope abundances of 92.2, 4.7 and 3.1\%, respectively \cite{IUPAC}. These isotopes can undergo the three main processes ((n,$\gamma$), (n,p) and (n,$\alpha$)) for neutron capture. Out of the nine possible outcomes, only five reach radioactive nuclides ($^{31}$Si, $^{28}$Al, $^{29}$Al, $^{30}$Al and $^{27}$Mg ). However, $^{30}$Al is hardly observable since its half-life is 3.6s and its cross section is only important above 8 MeV, almost at the end of the reactor's neutron spectrum. The cross sections for $^{30}$Si(n,$\alpha$)$^{27}$Mg is only relevant also at higher energies (above 6 MeV) where the neutron fluxes of the nuclear research reactor are lowest, thus very few photon counts are expected. In the case of $^{30}$Si(n,$\gamma$)$^{31}$Si, even if the cross section and the neutron flux are high at low energies, the gamma intensity of the related peak is very low (0.055\%), thus the expected number of counts is low and can be confused with a single escape peak of $^{28}$Al. Therefore only two radioactive nuclides, shown in Table \ref{tab:validation_results_Si_Ge}, have been observed: $^{28}$Al and $^{29}$Al.

Germanium has five natural isotopes: $^{70}$Ge, $^{72}$Ge, $^{73}$Ge, $^{74}$Ge and $^{76}$Ge, with average relative isotope abundances of 20.84, 27.54, 7.73, 36.28  and 7.61\%, respectively \cite{IUPAC}. These isotopes  can undergo neutron capture reactions via the same three main processes. There are fifteen possible outcomes, however, the reaction cross sections of (n,p) and (n,$\alpha$) are almost three orders of magnitude below the highest (n,$\gamma$) reaction, compared to just one order of magnitude difference in the case of silicon. Therefore all (n,p) and (n,$\alpha$) reactions are suppressed and only (n,$\gamma$) is favored. There are five such cases, of which only three lead to radioactive nuclides ($^{71}$Ge, $^{75}$Ge and $^{77}$Ge). From these, $^{71}$Ge is not observable since it has one decay mode (T$_{1/2}$=11.4 days) which only emits X-rays that cannot be detected with the gamma spectrometer. The other decay is metastable with T$_{1/2}$=20ms and even if it emits 174.9 keV photons, it has already disappeared by the time a spectroscopic measurement can be done. Thus only two radioactive nuclides, shown in Table \ref{tab:validation_results_Si_Ge}, are observed: $^{75}$Ge and $^{77}$Ge.

The expected produced radioactive nuclides described above are in agreement with the residual nuclei results from FLUKA shown in Table \ref{tab:resnuclei_results}. 

\begin{table}
\centering
    \begin{tabular}[t]{|c|c|c|c|}
   \hline 
\multicolumn{4}{|c|}{\textbf{Silicon products}} \\
\hline
A & 12 (Mg) & 13 (Al) & 14 (Si) \\ 
\hline
25	&	0.15	$\pm$	0.01	&				&				\\
26	&	0.02	$\pm$	0.001	&				&				\\
27	&	$\approx$0			&				&				\\
28	&				&	0.34	$\pm$	0.01	&	10.99	$\pm$	0.04	\\
29	&				&	0.012	$\pm$	0.001	&	100.00	$\pm$	0.06	\\
30	&				&	$\approx$0			&	3.91	$\pm$	0.01	\\
31	&				&				&	2.12	$\pm$	0.01	\\

\hline

  \hline
\multicolumn{4}{|c|}{\textbf{Germanium products}} \\
\hline
A & 30 (Zn) & 31 (Ga) & 32 (Ge) \\ 
\hline
67	&	0.0045	$\pm$	0.001	&				&		\\
69	&	$\approx$0			&				&				\\
70	&	0.0002	$\pm$	0.017	&	0.005	$\pm$	0.001	&	1.32	$\pm$	0.02	\\
71	&	$\approx$0			&				&	26.88	$\pm$	0.01	\\
72	&				&	0.001	$\pm$	0.0002	&	2.17	$\pm$	0.03	\\
73	&	$\approx$0			&	0.0002	$\pm$	0.0002	&	10.76	$\pm$	0.01	\\
74	&				&	$\approx$0			&	100.00	$\pm$	1.45	\\
75	&				&				&	7.96	$\pm$	0.02	\\
76	&				&	$\approx$0			&	0.69	$\pm$	0.01	\\
77	&				&				&	0.52	$\pm$	0.01	\\

\hline
    \end{tabular}%

\caption{Residual nuclei, by atomic number and mass number (A), estimated with FLUKA for the silicon and germanium samples. Values have been normalized to the maximum of each sample.}
\label{tab:resnuclei_results}
\end{table}

Some important comments are in order. We have neglected the metastable signatures and the peaks with gamma intensity lower than 1\%, in order to make a fair comparison with FLUKA. The simulation reproduces the metastable isotopes and their disappearance, which are characterized by a short half-life. Given the latter, the uncertainty in the measured time interval is very large. Thus we exclude the metastable decays ($^{74}$Ge(n,$\gamma$) $^{75m}$Ge, $t_{1/2}$=47.7s and $^{76}$Ge(n,$\gamma$)$^{77m}$Ge, $t_{1/2}=$ 53.7s). The gamma intensity cut is at the limit of detection capabilities, removing six peaks of $^{75}$Ge and two peaks of $^{77}$Ge. 

In addition, identified background peaks are not reported in these results. These peaks correspond to the K$_\alpha$ and K$_\beta$  characteristic X-rays emitted by lead from the detector shielding, a peak from electron-positron pair annihilation, peaks from double and single escapes, sum peaks and also $^{40}K$ natural background radiation. 

\begin{figure}[bt]
\begin{subfigure}{.33\textwidth}
  \centering
  \includegraphics[width=.99\linewidth]{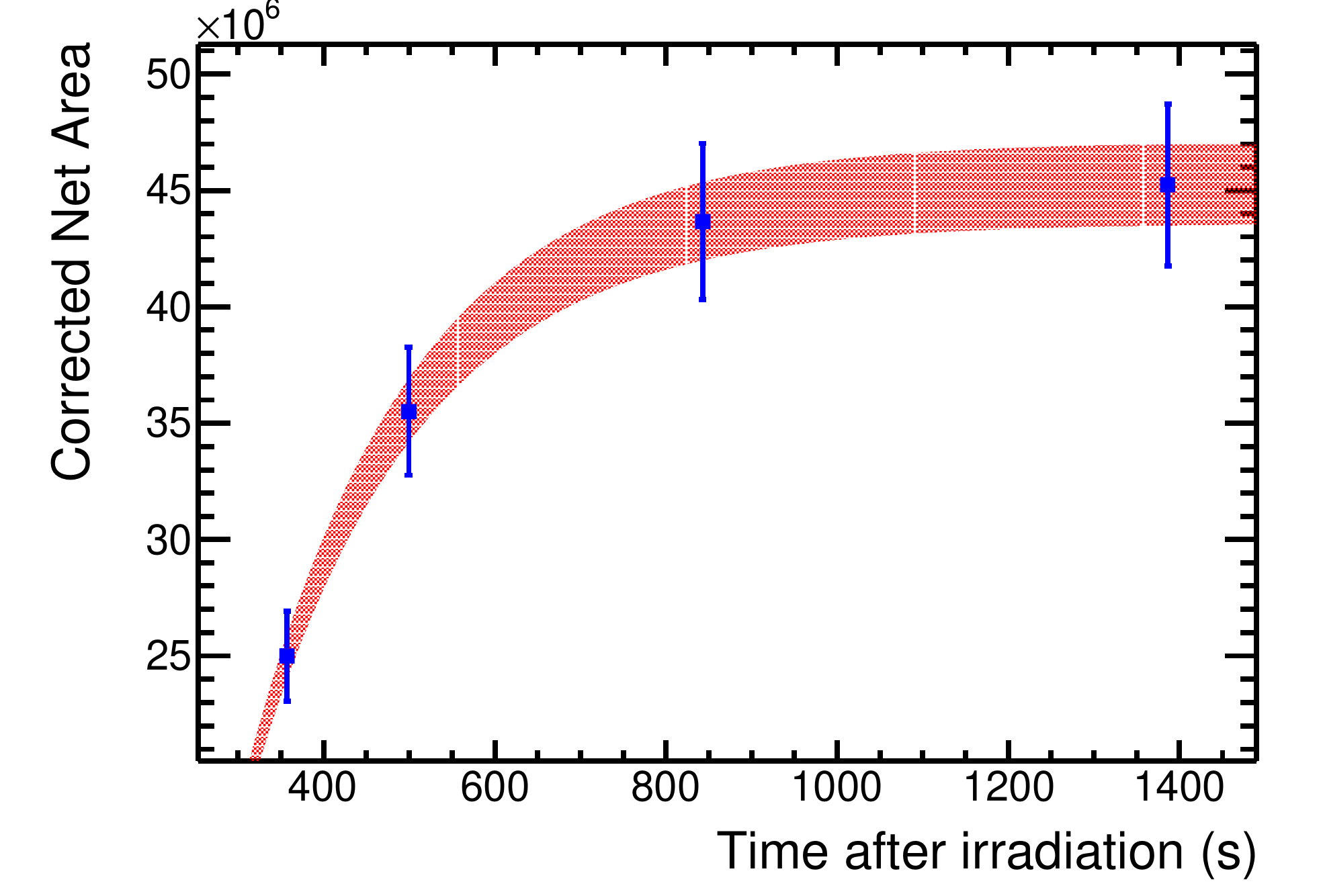}
  \caption{$^{28}$Al 1779 keV peak}
  \label{fig:timeevolAl28}
\end{subfigure}%
\begin{subfigure}{.33\textwidth}
  \centering
  \includegraphics[width=.99\linewidth]{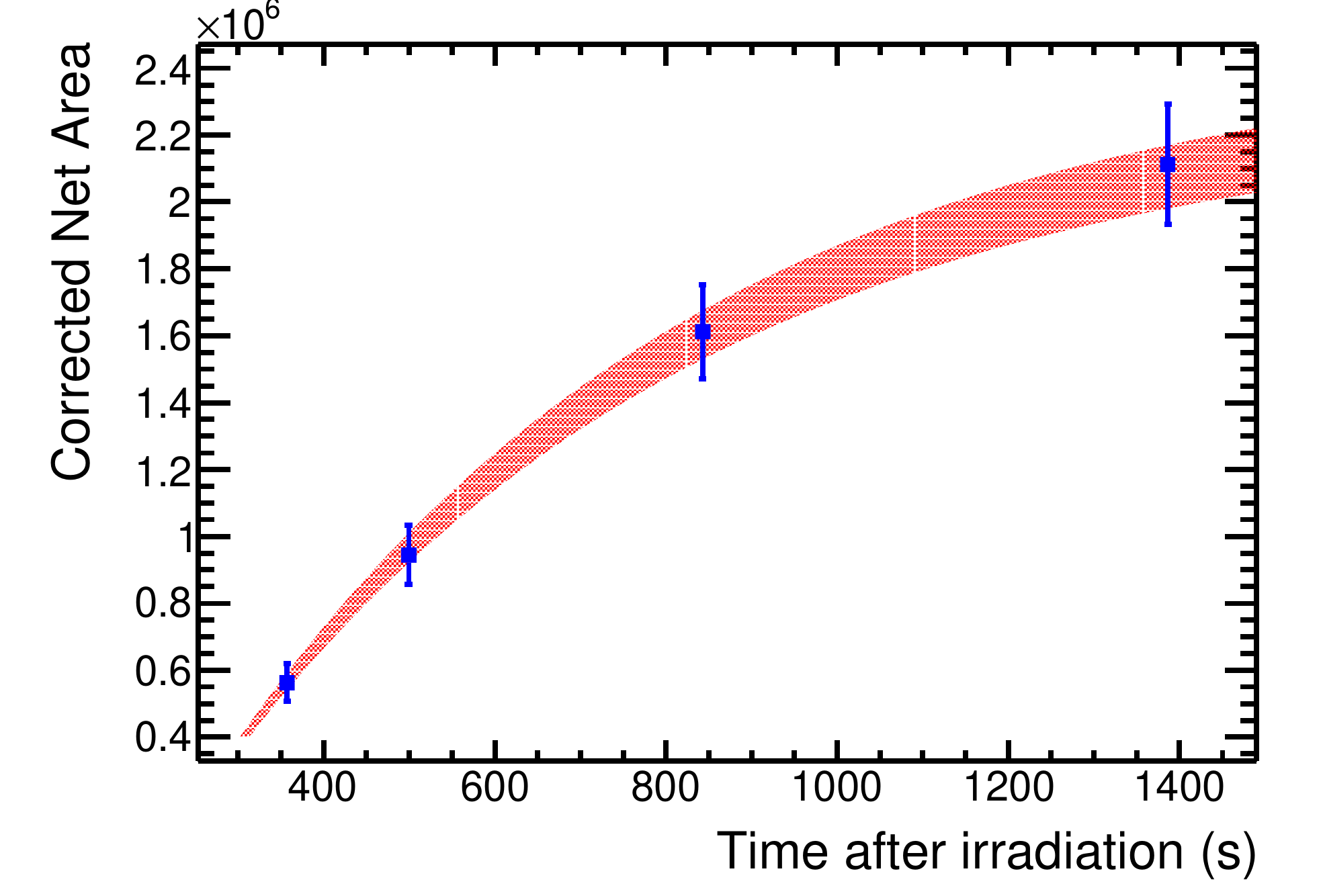}
  \caption{$^{29}$Al 1273 keV peak}
  \label{fig:timeevolAl29_1}
\end{subfigure}
\begin{subfigure}{.33\textwidth}
 \includegraphics[width=.99\linewidth]{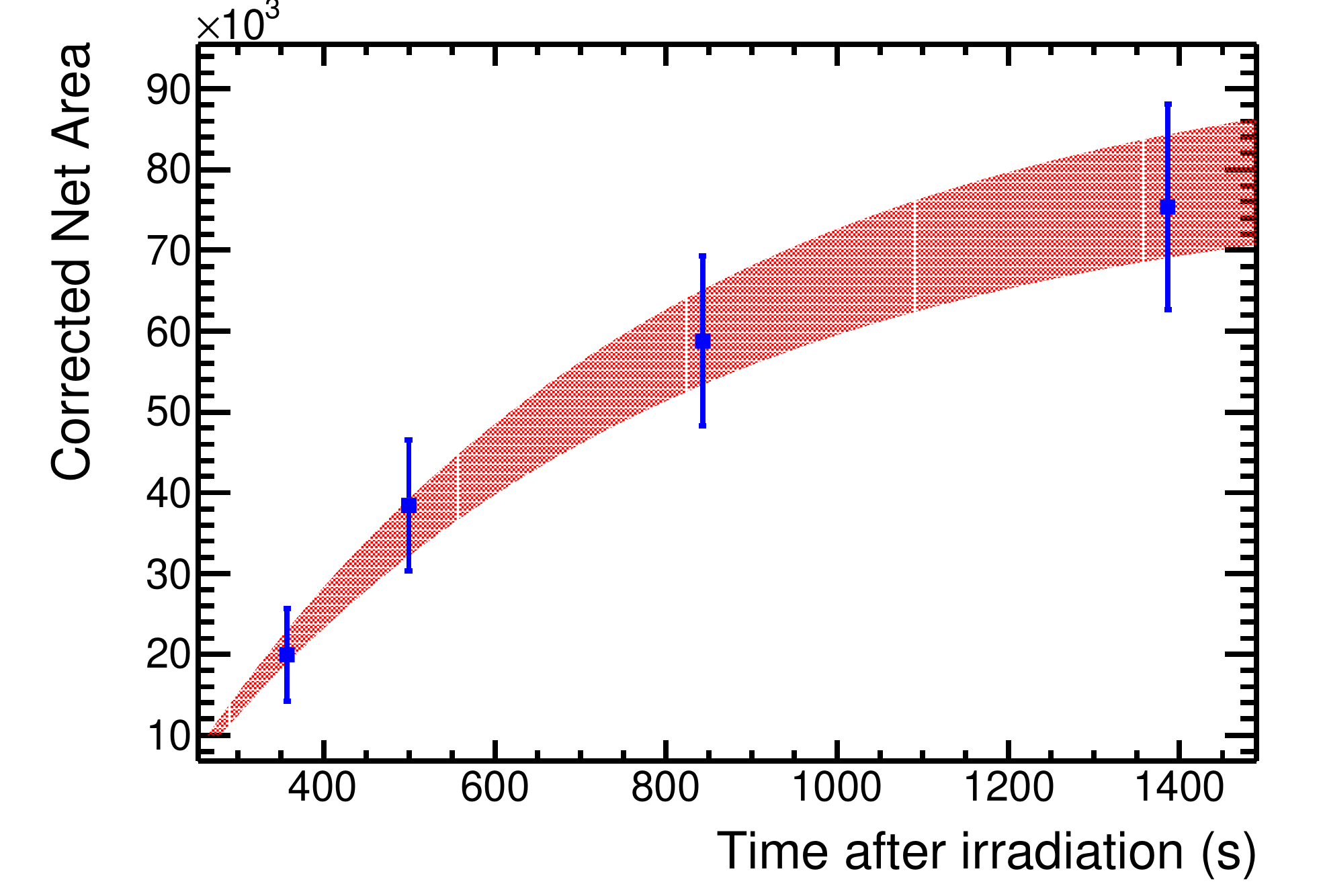}
  \caption{$^{29}$Al 2028 keV peak}
  \label{fig:timeevolAl29_2}
\end{subfigure}%
\caption{Time evolution of spectrometer measurements with theoretical fitting for silicon products. The shadowed band represents the error of the fit. Points are drawn at the end of the measurement and include the decay time.}
\label{fig:timeevol}
\end{figure}

In order to make a direct comparison between the FLUKA simulation results with those from the neutron irradiation measurements, we have corrected the effects of the spectrometer in the latter by dividing the net area of each peak by its efficiency at the given energy (see Fig. \ref{fig:Spect_eff}) and by the uptime fraction (see Table \ref{tab:samples_description}). The statistical and systematic errors on the measurements, displayed in Table \ref{tab:validation_results_Si_Ge}, are given by the uncertainties in the net area (counts) and in the detector efficiency, respectively. The uncertainties on the simulation are given by the statistical error of the simulation and the systematic error, found by
adding in quadrature the error on the neutron flux (shape and mean value, estimated to be $\approx 10$\%) and the normalization error in the photon peaks, that we explain below.

All FLUKA photon peak amplitudes are normalized by a single constant for each sample that minimizes the difference between simulation and experimental measurement, thus only relative peak amplitudes will be compared.

This procedure is done as follows: first, for each energy peak, the time evolution of data during the four measurements at different times is fitted with the following function:
\begin{equation}
N_{0} \left(e^{-\frac{t_{0}}{\tau}}-e^{-\frac{t}{\tau }} \right)
\label{Eq_timeevol}
\end{equation}
where $N_{0}$ is the starting number of radionuclides, $t_0$ is the starting time of the measurement and $\tau$ is the lifetime of the radionuclide. As an example of these fits, we  show in Fig. \ref{fig:timeevol} the results for the silicon sample. 

From the fit to the experimental data we extract $N_{0}^{data}$, which will be compared with the one obtained using the FLUKA simulation, $N_0^{sim}$. Assuming no variation in the nuclides natural abundances, the ratio $N_0^{data}/N_0^{sim}$ should be a constant for all energy peaks of the same radionuclide. Thus we fit the ratios $N_0^{data}/N_0^{sim}$ for all peaks with a constant. We use this constant as global normalization for the FLUKA photon peaks amplitudes. The error associated with the constant fit is estimated to be 0.84\% and 0.17\% for the silicon and germanium samples, respectively.

\begin{figure}[th]
\begin{subfigure}{.5\textwidth}
  \centering
  \includegraphics[width=.99\linewidth]{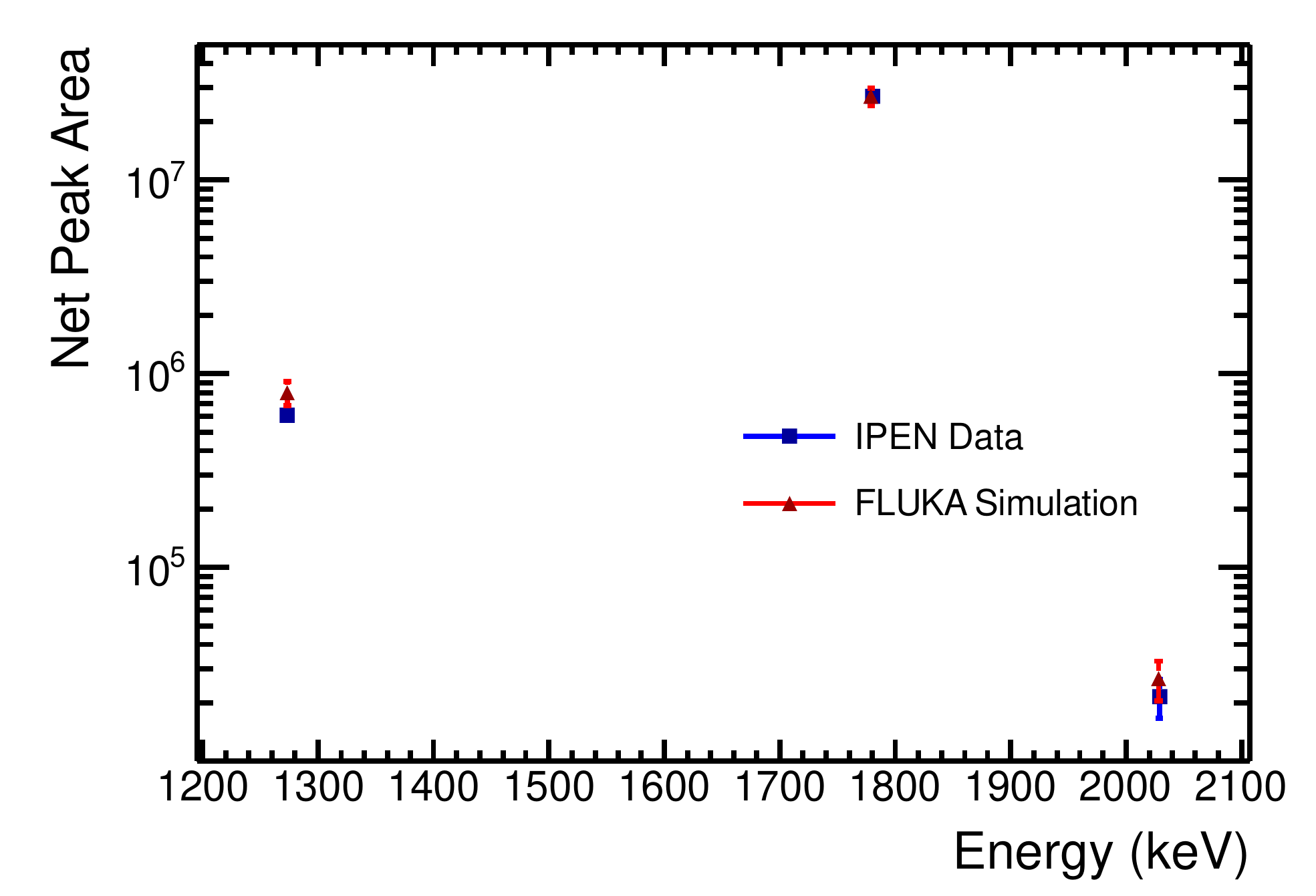}
  \caption{Silicon sample}
\label{fig:ipen_fluka_Si_peaks}
\end{subfigure}%
\begin{subfigure}{.5\textwidth}
  \centering
  \includegraphics[width=.99\linewidth]{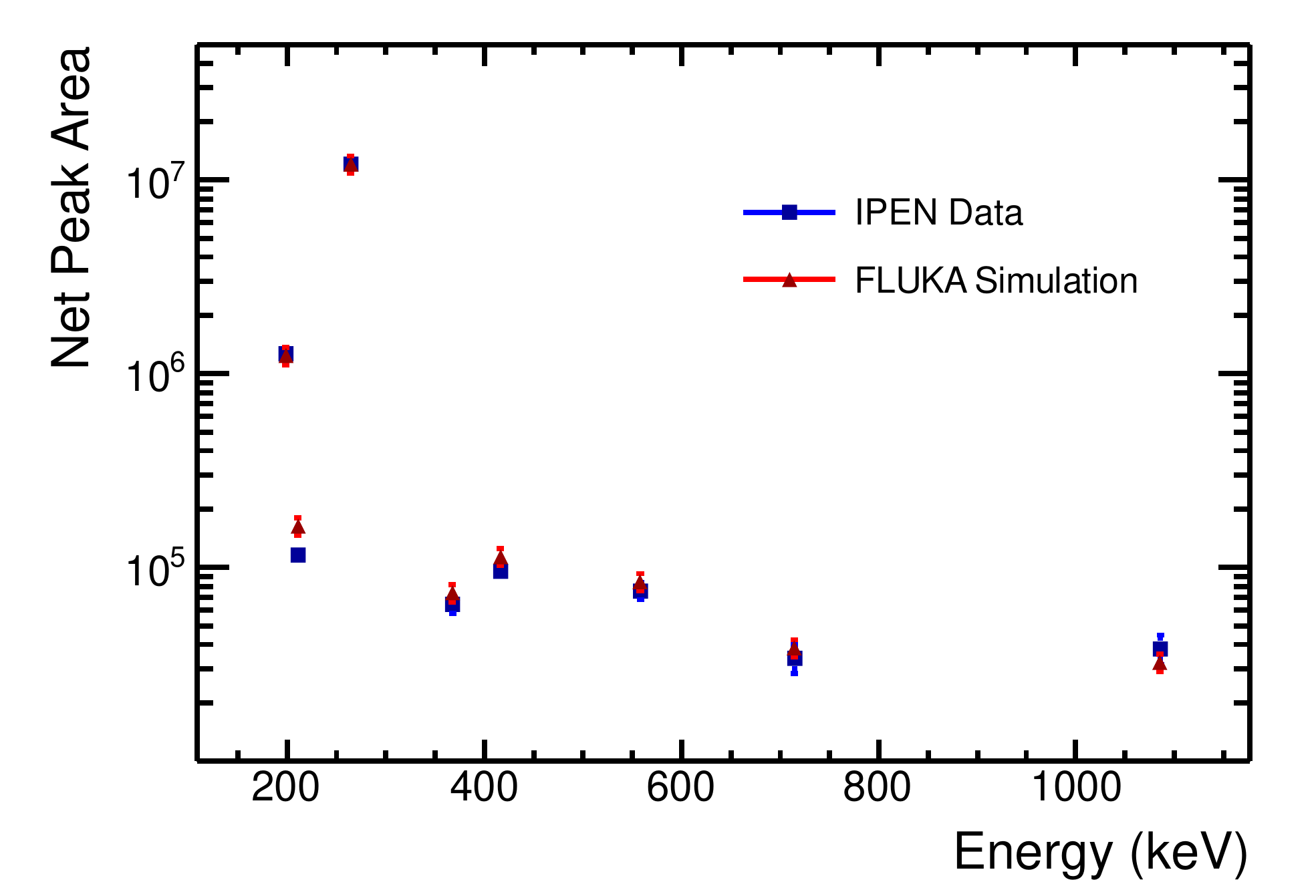}
  \caption{Germanium sample}
   \label{fig:ipen_fluka_Ge_peaks}
\end{subfigure}
\caption{Gamma spectrum of neutron activated samples comparing experimental measurements (IPEN Data)and FLUKA simulations for peaks with intensities greater than 1\% and non metastable decays. The Si sample was irradiated for 60 s at 4.5 MW and the counts collected for 168.33 s after 202 s decay time. The Ge sample was irradiated for 20 s at 4.5 MW and the counts collected for 205.53 s after 235 s decay time.}
\label{fig:peaks_comparison}
\end{figure}

In Fig. \ref{fig:ipen_fluka_Si_peaks} and  \ref{fig:ipen_fluka_Ge_peaks} the gamma spectrum of neutron-activated silicon and germanium comparing experimental data with FLUKA simulation is presented. We show the results only for the first measurement in time as a benchmark, since it has the largest error, due to the lower number of counts.  

The full results, including the capture reaction, half-life, peak energy and gamma intensity, are given in Table \ref{tab:validation_results_Si_Ge} where the difference between data and simulation is given in terms of sigmas ($\frac{\left|N_{exp}-N_{sim}\right|}{\sqrt{\sigma_{exp}^2+\sigma_{sim}^2}}$).  The average difference for both materials in photon peak amplitude is 0.85 $\sigma$. We also observe that FLUKA is capable of predicting relative peak amplitudes of produced radionuclides with an average uncertainty of 12.9\%, evaluated with $\frac{|N_{exp}-N_{sim}|}{N_{sim}}\times100\%$. The average total error on the measurements is 10.4\% and on the simulations 11.7\%, while for the simulation it is mostly driven by the error on the neutron flux, for the experimental data it is due to the uncertainty in the detector efficiency and the statistical error of the counts.  

\begin{landscape}
\begin{longtable}{|c|c|c|c|c|c|c|c|}
\hline 
Nuclide &  Half-life&IAEA  & Measured & IAEA & Corrected Net  & FLUKA &  $\sigma$\\
& (min) &energy (keV)& energy (keV)& intensity &Peak Area$\dagger$ & simulation* & \\

\hline 
 $^{28}$Si(n,p) $^{28}$Al & 2.24 &1778.99& 1780.1$\pm$2.0 & 100& (2.71$\pm0.01\pm0.19) \times 10^7$  & (2.70$\pm0.04\pm0.27) \times 10^7$ & 0.06\\
$^{29}$Si(n,p) $^{29}$Al & 6.56 &1273.36 & 1273.2$\pm$1.8 & 91.3& (6.12$\pm0.18\pm0.41) \times 10^5$   &(7.97$\pm0.76\pm0.78) \times 10^5$ &1.56\\
 &  &2028.1& 2029.1$\pm$2.1 & 3.5& (2.16$\pm0.46\pm0.15 )\times 10^4$  & (2.67$\pm0.55\pm0.27) \times 10^4$& 0.65 \\

\hline

$^{74}$Ge(n,$\gamma$) $^{75}$Ge & 82.78 &  198.6& 198.9$\pm$1.1 & 1.19 &  (1.27$\pm0.02\pm0.04) \times10^6$  &(1.24$\pm0.01\pm0.12) \times 10^6$ & 0.22\\

$^{76}$Ge(n,$\gamma$) $^{77}$Ge & 672.6& 211.03&211.3$\pm$1.1 & 30 & (1.16$\pm0.09\pm0.04) \times10^5$ & (1.63$\pm0.02\pm0.16) \times 10^5$ & 2.48 \\

& & 367.49& 367.7$\pm$1.2 & 14.5 & (6.46$\pm0.65\pm0.22)\times10^4$ & (7.38$\pm0.14\pm0.74) \times 10^4$& 0.91\\

&  & 416.35& 416.7$\pm$1.3 & 22.7 & (9.57$\pm0.71\pm0.33)\times10^4$ &(1.13$\pm0.02\pm0.11) \times 10^5$ & 1.28\\

&  & 557.92& 558.4$\pm$1.4 & 16.8 & (7.57$\pm0.67\pm0.31) \times10^4$ & (8.43$\pm0.16\pm0.84) \times 10^4$ & 0.74\\

&  & 714.37& 714.7$\pm$1.5 & 7.5 & (3.39$\pm0.56\pm0.13) \times10^4$ & (3.85$\pm0.09\pm0.39) \times 10^4$& 0.65\\

&  & 1085.23& 1085.8$\pm$1.7 & 6.4 & (3.80$\pm0.65\pm0.14) \times10^4$ &(3.23 $\pm0.07\pm0.32) \times 10^4$ & 0.78\\
\hline
Combined peaks& &   &    &  &  & & \\
$^{75}$Ge+$^{77}$Ge	 & & 264.6/264.5 & 264.9$\pm$1.1 & 11.4/53.3 & (1.21$\pm0.01\pm0.04) \times10^7$ &(1.21$\pm0.004\pm0.12) \times 10^7$ &  0.02\\
\hline
 
\caption{Validation results for Si and Ge comparing experimental measurements (ORTEC spectrometer model GEM70P4) and FLUKA simulations for the main gamma peaks from nuclide decays. $\dagger$Corrected for efficiency and uptime. *FLUKA simulation results are normalized as described before. Both experimental and simulation results include the statistical and systematic errors, as described in the text. The last column represents the difference between experiment and simulation as a number of standard deviations: $\frac{\left|N_{exp}-N_{sim}\right|}{\sqrt{\sigma_{exp}^2+\sigma_{sim}^2}}$ .}
\label{tab:validation_results_Si_Ge}
\end{longtable}
\end{landscape}

\section{Conclusions}

FLUKA has been confirmed to give reliable results for neutron activation of two semiconductor materials, silicon and germanium, in reactor experiments. Since the nuclear research reactor energies are below 10 MeV, we have tested FLUKA's low energy neutron library (ENDF/B-VIIR) that predicts the corresponding cross sections for all interactions. In addition, the decay photon scoring reproduces the expected counts in the gamma spectrometer. It should be noted that FLUKA has a limitation to simulate the whole transmutation of the sample since it keeps the target material constant during the simulation. 

The gamma emitted spectrum of radionuclides produced by these materials irradiated at a nuclear research reactor has been measured using a high purity Ge gamma spectrometer. These results have been compared to the predictions from FLUKA obtaining a mean agreement within 13\% for the intermediate-lived radioisotopes and peaks with gamma intensities greater than 1\%.

Simulating with high reliability the neutron transmutation of semiconductors is useful in doping studies, as well as in evaluating the radiation damage, since part of it is due to the change in the chemical composition of the active sensor, altering its properties. 


\section{Acknowledgments}

The authors gratefully acknowledge DGI-PUCP for financial support under Grants No. DGI-2015-192 and DGI-2017-3-0019, as well as CONCYTEC under Grant No. FONDECYT-2013-102. The authors wish to thank the FLUKA scientific committee for useful comments and suggestions, Jose Fernandes and Francisco Rumiche for the SEM measurements and Jorge Andr\'es Guerra, Patrizia Pereyra and Daniel Palacios for useful discussions. 

\section*{References}
\bibliography{SiGebibfile}

\end{document}